\documentclass[11pt]{article}
\begin{document}
\title{{\bf{\Large  Noncommutativity in interpolating string: A study of gauge
 symmetries in noncommutative framework}}}
\author{
{\bf {\normalsize Sunandan Gan{g}opadhyay}$^{a,}$\thanks{sunandan@bose.res.in}},\,\,{\bf {\normalsize Arindam Ghosh Hazra}$^{a,}$\thanks{arindamg@bose.res.in}}
\\{\bf {\normalsize Anirban Saha}$^{b,}$\thanks{ani\_saha09@dataone.in}}\\
$^{a}$ {\normalsize S.~N.~Bose National Centre for Basic Sciences,}\\
{\normalsize JD Block, Sector III, Salt Lake, Kolkata-700098, India}\\[0.3cm]
$^{b}$ {\normalsize Department of Physics, Presidency College,}
\\{\normalsize 86/1 College Street, Kolkata-700073, India.}\\[0.3cm]
}

\maketitle

\centerline{\large \bf Abstract}
A new Lagrangian description that interpolates between the 
Nambu--Goto and Polyakov version of interacting strings is given. 
Certain essential modifications in the Poission bracket structure 
of this interpolating theory generates noncommutativity among 
the string coordinates for both free and interacting strings. 
The noncommutativity is shown to be a direct consequence
of the nontrivial boundary conditions. A thorough analysis of the 
gauge symmetry is presented taking into account the new modified 
constraint algebra, which follows from the noncommutative structures
and finally a smooth correspondence between gauge symmetry and reparametrisation is established.

\vskip 0.2cm
{\bf PACS} 11.15.-q, 11.10 Nx, 11.25.-w \\
\vskip 0.3cm
{\bf Keywords: } Strings, Noncommutativity, Gauge symmetry\\

\section{Introduction}
For the last few decades string theory has been regarded as
the most promising step towards the fundamental theory uniting all the basic 
interactions at the Planck scale \cite{pol}. The dynamics of a
 bosonic string is described either by Nambu--Goto (NG) or 
Polyakov action. Both these actions, though very well-known 
in the literature, poses certain degrees of difficulty. 
NG formalism is inconvenient for path integral quantisation whereas Polyakov 
action involves many redundant degrees of freedom. However, 
yet another formulation, interpolating between these two versions 
of string action, has also been put forward in the literature  
\cite{rb}. This interpolating Lagrangian is a better description of 
the theory in the sense that it neither objects to quantization 
nor has as many redundancies as in the Polyakov version. 
Further, it gives a perfect platform to study the gauge 
symmetries vis-\`{a}-vis reparametrisation symmetries of 
the various free string actions by a constrained Hamiltonian 
approach \cite{rbpmas1, rbpmas2}. 

Noncommutative (NC) theories, on the other hand has been one
of the central areas of research recently \cite {sz}. 
In this context the
study of open strings propagating in the presence of a 
background Neveu--Schwarz two form field $B_{\mu\nu}$ has become 
important because it exhibits a manifest NC
structure among the space-time coordinates of the D-branes
\cite {sw}. Several approaches have been taken to obtain 
such results, for example a Dirac approach \cite{dir} is employed 
with the string boundary conditions (BC(s)) imposed as  
second class constraints in \cite{ar, br}.
However, in a series of recent papers 
\cite{rb, rbbckk, bcsgaghs}, it has been shown explicitly
that noncommutativity can be obtained in a more transparent way
by modifying the canonical Poisson bracket (PB) structure, so that
it is compatible with the BC(s). This is similar in spirit to the 
treatment of Hanson, Regge and 
Teitelboim \cite{hrt}, where modified PB(s) were
obtained for the free NG string. In a very recent paper \cite{agh}, it has
also been obtained using a conformal field theoretic approach.

In the present paper, acknowledging the above facts, we derive 
a master action for interacting bosonic strings, 
interpolating between the NG and Polyakov formalism. 
Modification of the basic PB structure compatible with BC(s) followed by 
the emergence of the noncommutativity is shown
in this formalism (in case of both free and interacting strings)
 following the approach in \cite{rb, rbbckk, bcsgaghs}. 
Our results go over smoothly to the Polyakov version once
proper identifications are made. Interestingly, we observe
that a gauge fixing is necessary to give an exact NC solution between
the string coordinates. 
Further, this gauge fixing condition restrict us to 
a reduced phase space of the interpolating theory\footnote{As far as the study
of gauge symmetry
is concerned, we consider only free strings in the 
remainder of the paper.} which in turn 
minimizes the gauge redundancy 
of the theory by identifying a particular combination of the 
constraints (that occurs in the full gauge independent theory) leading
to a new involutive constraint algebra which is 
markedly different from that given in \cite{rb}.  
With the above results at our disposal, we go over to the
study of gauge symmetry in the NC framework.
Owing to the new constraint algebra we find
surprising changes in the structure constants of the theory.
Finally, we compute the gauge variations of the fields and show
the underlying unity of diffeomorphism with the gauge symmetry
in the NC framework. 

The organisation of the paper is as follows. In the 
next section we briefly review interacting string in NG 
formalism. This fixes the notations. Here we also extend 
the domain of definition of the fields and give the closure 
relations of Virasoro algebra. 
In section 3 we formulate the interpolating Lagrangian for 
interacting string, derive their corresponding boundary 
conditions and full set of constraints. Their identification 
with the usual NG and Polyakov
 version is demonstrated. Section 4 elaborates on the 
modification of the basic PB(s) of the 
interpolating theory (for both free and interacting case) 
to make the canonical structure compatible with the BC(s). 
In this section, one finds the emergence of noncommutative behavior 
among the coordinates. To give an explicit NC solution
one needs to put  a restriction on the phase space of the 
interpolating theory. In section 5 we 
systematically analyse the reduced gauge 
symmetry of bosonic string in light of the modified 
canonical setup.
Finally we conclude in section 6.

\section{Interacting  String in Nambu-Goto Formalism}

In this section, we analyse the NG  formulation of the
interacting bosonic string. As we shall see in the next 
section, this is essential in the construction of 
the Interpolating Lagrangian for the interacting string.
The NG action for a bosonic string moving in the presence
of a constant background Neveu-Schwarz two-form field $B_{\mu\nu}$
reads:
\begin{eqnarray}
S_{NG}=\int^{\infty}_{- \infty} d\tau \int^{\pi}_{0}d\sigma 
\left[\mathcal L_{0}+eB_{\mu\nu}\dot{X}^\mu
X^{\prime \nu}\right]
\label{NGaction} 
\end{eqnarray}
where $\mathcal L_{0}$ is the free NG Lagrangian density
given by\footnote{Here $X^{\prime \mu} = \frac{\partial 
X^{\mu}}{\partial \sigma}$
and  $\dot{X}^\mu = \frac{\partial X^{\mu}}{\partial \tau}$.}
\begin{eqnarray}
\mathcal L_{0}=-\left[(\dot{X}.X^{\prime})^2-\dot{X}^{2}X^{\prime 2}\right]
^\frac{1}{2} .
\label{freeNGaction} 
\end{eqnarray}
The string tension is kept implicit for convenience.
The Euler-Lagrange (EL) equations and BC
obtained by varying the action read:
\begin{eqnarray}
\dot{\Pi}^{\mu} + K^{\prime \mu} &=& 0 \nonumber \\
K^{\mu}\vert_{\sigma = 0, \pi} &=& 0
\label{EL} 
\end{eqnarray}
where,
\begin{eqnarray}
\Pi_{\mu} &=& \frac{\partial {\mathcal L}}{\partial \dot{X}^{\mu}}
= \mathcal L_{0}^{-1}\left(-X^{\prime 2}\dot{X}_{\mu}
+(\dot{X}.X^{\prime})X^{\prime}_{\mu}\right) + eB_{\mu\nu}X^{\prime \nu}
\nonumber \\
K_{\mu} &=& \frac{\partial {\mathcal L}}{\partial X^{\prime \mu}}
= \mathcal L_{0}^{-1}\left(- \dot{X}^2 X^{\prime}_{\mu}
+ (\dot{X} \cdot X^{\prime})\dot{X}_{\mu}\right) -e B_{\mu\nu}\dot{X}^{\nu}.
\label{NGmomentum} 
\end{eqnarray}
Note that $\Pi_{\mu}$ is the canonically conjugate momentum
to $X^{\mu}$.
The nontrivial PB(s) of the theory are given by:
\begin{eqnarray}
\{X^{\mu}(\tau,\sigma), \Pi^{\nu}(\tau,\sigma^{\prime})\}=\eta^{\mu \nu}
\delta(\sigma-\sigma^{\prime}).
\label{NGPB} 
\end{eqnarray}
The primary constraints of the theory are:
\begin{eqnarray}
\Omega_{1} &=& \Pi_{\mu}X^{\prime \mu}\approx 0\nonumber\\
\Omega_{2} &=& \left(\Pi_{\mu} - e B_{\mu \nu}
X^{\prime \nu}\right)^2 + X^{\prime 2}
\approx 0.
\label{NGconstraints} 
\end{eqnarray}
From the above PB structure (\ref{NGPB}), it is easy to 
generate a first class (involutive) algebra:
\begin{eqnarray}
\{\Omega_{1}(\sigma), \Omega_{1}(\sigma^{\prime})\}
&=& \left[\Omega_{1}(\sigma)+\Omega_{1}(\sigma^{\prime})\right]
\partial_{\sigma}\delta(\sigma-\sigma^{\prime}) \nonumber \\
\{\Omega_{1}(\sigma), \Omega_{2}(\sigma^{\prime})\}
&=& \left[\Omega_{2}(\sigma)+\Omega_{2}(\sigma^{\prime})\right]
\partial_{\sigma}\delta(\sigma-\sigma^{\prime})\nonumber \\
\{\Omega_{2}(\sigma), \Omega_{2}(\sigma^{\prime})\}
&=& 4\left[\Omega_{1}(\sigma)+\Omega_{1}(\sigma^{\prime})\right]
\partial_{\sigma}\delta(\sigma-\sigma^{\prime}).
\label{NGalgebra00} 
\end{eqnarray}
Now as happens for a reparametrisation invariant theory,
 the canonical Hamiltonian density defined by a Legendre transform vanishes
\begin{eqnarray}
\mathcal H_{c} = \Pi_{\mu}\dot{X}^{\mu}-\mathcal L = 0.
\label{CanonicalH1} 
\end{eqnarray}
This can be easily seen by substituting 
(\ref{NGmomentum}) in (\ref{CanonicalH1}).
The total Hamiltonian density is thus given by 
a linear combination of the first class
constraints (\ref{NGconstraints}):
\begin{eqnarray}
\mathcal H_{T}=-\rho\Omega_{1}-\frac{\lambda}{2}\Omega_{2}
\label{totalH} 
\end{eqnarray}
where $\rho$ and $\lambda$ are Lagrange multipliers. It is easy
to check that time preserving the primary constraints yields no
new secondary constraints. Hence the total set of constraints
 of the interacting NG theory is given by the 
first-class system (\ref{NGconstraints}).\\
\noindent
Now we enlarge the domain
of definition of the bosonic field $X^{\mu}$ from $[0,\pi]$
to $[-\pi,\pi]$ by defining \cite{bcsgaghs}{\footnote{
This is done in order to write down the generators of
$\tau$ and $\sigma$ reparametrisation in a compact form.}}
\begin{eqnarray}
X^{\mu}(\tau , -\sigma) = X^{\mu}(\tau , \sigma)\ \ ; \ \  
B_{\mu \nu} \to - B_{\mu \nu} \ \mathrm{under}\ \sigma \to -\sigma.
\label{37}
\end{eqnarray}
The second condition implies that $B_{\mu \nu}$, albeit a constant, 
transforms as a pseudo scalar under $\sigma \to -\sigma$ 
in the extended interval. This ensures that the interaction term
$eB_{\mu\nu}\dot{X}^{\mu}X^{\prime \nu}$ in (\ref{NGaction})
remains invariant under $\sigma \to - \sigma$ like the
free NG Lagrangian density ${\cal{L}}_{0}$ (\ref{freeNGaction}).
Consistent with this, we have
\begin{eqnarray}
\Pi^{\mu}(\tau , -\sigma) = \Pi^{\mu}(\tau , \sigma),
\quad 
X^{\prime \mu}(\tau , -\sigma) = - X^{\prime \mu}(\tau , \sigma). 
\label{37h}
\end{eqnarray}
Now, from  (\ref{NGconstraints}), (\ref{37}) we note that
the constraints $\Omega_1(\sigma) \approx 0$
and $\Omega_2(\sigma) \approx 0$ are odd and even respectively under
$\sigma \rightarrow -\sigma$. 
Now demanding the total Hamiltonian density $\mathcal{H_T}$
(\ref{totalH}) also remains invariant under $\sigma
\to - \sigma$, one finds that $\rho$ and $\lambda$ must be
odd and even respectively under $\sigma \to - \sigma$.

\noindent
We may then write the generator of all $\tau$ and $\sigma$ 
reparametrisation as the functional \cite{hrt}:
\begin{eqnarray}
L[f] = \frac{1}{2}\int_{0}^{\pi} d\sigma \{f_+(\sigma) \Omega_2(\sigma)
+ 2 f_-(\sigma) \Omega_1(\sigma)\}\,,
\label{32u}
\end{eqnarray}
where, $f_{\pm}(\sigma)=\frac{1}{2}(f(\sigma)\pm f(-\sigma))$ are by
construction even and odd function and  $f(\sigma)$ is an arbitary
differentiable function defined in the extended interval $[-\pi, \pi]$.
The above expression can be simplified to:
\begin{eqnarray}
L[f] &=& \frac{1}{4}\int_{-\pi}^{\pi} d\sigma f(\sigma)
\left[\Omega_2(\sigma) + 2\Omega_1(\sigma)\right] \nonumber \\
&=& \frac{1}{4}\int_{-\pi}^{\pi} d\sigma f(\sigma)
\left[\Pi_{\mu}(\sigma) + X^{\prime}_{\mu}(\sigma) 
-e B_{\mu \nu}X^{\prime \nu}(\sigma) \right]^2
\label{33u}.
\end{eqnarray}
It is now easy to verify (using (\ref{NGalgebra00})) that 
the above functional (\ref{33u}) generates the following Virasoro algebra:
\begin{eqnarray}
\{L[f(\sigma)] , L[g(\sigma)]\} &=& L[f(\sigma)g^{\prime}(\sigma)
- f^{\prime}(\sigma)g(\sigma)].
\label{34a}
\end{eqnarray}
Defining 
\begin{eqnarray}
L_m = L[e^{-im\sigma}]\,,
\label{34y}
\end{eqnarray}
one can write down an equivalent form of the Virasoro algebra
\begin{eqnarray}
\{L_m , L_n\} &=& i (m-n) L_{m + n}\,.
\label{34b}
\end{eqnarray}
Note that we do not have a central extension here, as 
the analysis is entirely classical.

\section{Interpolating Lagrangian, boundary conditions
 and constraint structure of the Interacting String}
In the previous section we have reviewed the salient 
features of the interacting NG string.
We now pass on to the construction of the interpolating action of the
interacting string\footnote{The construction of the interpolating
action for the free string has been discussed in \cite{rb}.}. To 
achieve this end, we write down the Lagrangian
 of the interacting NG action in the
first-order form:
\begin{eqnarray}
\mathcal L_{I}=\Pi_{\mu}\dot{X}^{\mu}-\mathcal H_{T}\,.
\label{1stL} 
\end{eqnarray}
Substituting (\ref{totalH}) in (\ref{1stL}), $\mathcal L_{I}$ becomes
\begin{eqnarray}
\mathcal L_{I}=\Pi_{\mu}\dot{X}^{\mu}+\rho\Pi_{\mu}X^{\prime\mu}
+\frac{\lambda}{2}\left[(\Pi^2 + X^{\prime 2})-2eB_{\mu\nu}
\Pi^{\mu}X^{\prime\nu}+e^2B_{\mu\nu}B^{\mu}_{\ \rho}X^{\prime\nu}
X^{\prime\rho}\right]\,.
\label{subL} 
\end{eqnarray}
The advantage of working with the interpolating action is that
it naturally leads to either the NG or the Polyakov formulations 
of the string. In the Lagrangian (\ref{subL}), $\lambda$ and  $\rho$
originally introduced as Lagrange multipliers, will be treated as
independent fields, which behave as scalar and pseudo-scalar
fields respectively in the extended world-sheet, as was
discussed in the previous section.
 We will eliminate $\Pi_{\mu}$ from (\ref{subL}) as
it is an auxilliary field. The EL equation for 
 $\Pi_{\mu}$ is:
\begin{eqnarray}
\dot{X}^{\mu}+\rho X^{\prime\mu}+\lambda\Pi^{\mu}-
e\lambda B^{\mu\nu}X^{\prime}_{\nu}=0\,.
\label{momentumeqn} 
\end{eqnarray}
Substituting $\Pi_{\mu}$ from (\ref{momentumeqn})back in (\ref{subL}) yields:
\begin{eqnarray}
\mathcal L_{I}=-\frac{1}{2\lambda}\left[\dot{X}^{2}+
2\rho(\dot{X}.X^{\prime})+(\rho^2-\lambda^2)X^{\prime 2}-2\lambda eB_{\mu\nu}
\dot{X}^{\mu}X^{\prime\nu}\right]\,.
\label{interpolatingL} 
\end{eqnarray}
This is the form of the interpolating Lagrangian of the
interacting string.

The reproduction of the NG action (\ref{NGaction}) from the interpolating
action of the interacting string is trivial and can be done
by eliminating $\rho$ and $\lambda$ from their respective EL
equations of motion following from (\ref{interpolatingL}),
\begin{eqnarray}
\rho &=& - \frac{\dot{X} \cdot X^{\prime}}{X^{\prime 2}}\nonumber \\
\lambda^{2} &=& \frac{\left(\dot{X} \cdot X^{\prime}\right)^2 
- \dot{X}^2 X^{\prime 2}}{X^{\prime 2} X^{\prime 2}}.
\label{rholambdaeqn} 
\end{eqnarray}
If, on the other hand, we identify $\rho$ and $\lambda$ 
with the following contravariant
components of the world-sheet metric,
\begin{eqnarray}
g^{ab}=(-g)^{-{1\over 2}}\pmatrix{{1\over \lambda} & {\rho \over 
\lambda}\cr {\rho \over \lambda} &
{(\rho^2-\lambda^2)\over {\lambda}}}
\label{idm} 
\end{eqnarray}
then the above Lagrangian (\ref{interpolatingL})
reduces to the Polyakov form,
\begin{eqnarray}
\mathcal L_P= -{1\over 2}\left( {\sqrt {-g}}g^{ab}{\partial}_a
X^{\mu}{\partial}_bX_{\mu}
-e\epsilon^{ab}B_{\mu \nu}\partial_{a} X^{\mu}\partial_{b} X^{\nu}\right)
\ \ ; \  \ \left(a,b = \tau,\sigma\right).
\label{Polyaction} 
\end{eqnarray}
\noindent In this sense, therefore, the Lagrangian in (\ref{interpolatingL})
is referred to as an interpolating Lagrangian\footnote{It should be noted
that the interpolating action has only two additional degrees of
freedom, $\lambda$ and $\rho$, which does not fully match the three
degrees of freedom of the worldsheet metric of the Polyakov action.However,
due to Weyl
invariance of the Polyakov action, only two of the three different metric
coefficients $g_{ab}$ are really independent. This Weyl invariance
is special to the Polyakov string, the higher branes do not share it.}.

\noindent
We can now, likewise construct the interpolating BC
from the interpolating Lagrangian (\ref{interpolatingL}), 
\begin{eqnarray}
K^{\mu} = \left[\partial {\cal{L}}_{I}\over \partial 
X^{\prime}_{\mu}\right]_{\sigma =0,\pi} = \left({\rho \over \lambda}
{\dot X}^{\mu}+{\rho^2 - 
\lambda^2 \over \lambda}X'^{\mu} + eB^{\mu}_{\ \nu}\dot{X}^{\nu}
\right)_{\sigma = 0, \pi} = 0. 
\label{16} 
\end{eqnarray}
The fact that this can be easily interpreted as interpolating
BC, can be easily seen by using the expressions 
(\ref{rholambdaeqn}) for $\rho$ and $\lambda$ in  (\ref{16})
to yield:
\begin{eqnarray}
\left[\mathcal L^{- 1}_{0}\left( - \dot{X}^{2} X^{\prime \mu}
+ \left(\dot{X} X^{\prime}\right)\dot{X}^{\mu}\right) - 
e B^{\mu \nu}\dot{X}_{\nu}\right]_{\sigma = 0, \pi} = 0\,.
\label{17} 
\end{eqnarray}
This is the BC of the interacting NG string (\ref{NGmomentum}).

\noindent
On the otherhand, we can identify $\rho$ and $\lambda$ with the metric
components as in (\ref{idm}) to recast (\ref{16}) as: 
\begin{eqnarray}
\left(g^{1a}\partial_{a}X^{\mu}(\sigma) +
\frac{1}{\sqrt{-g}}e B^{\mu}_{\ \nu}\partial_{0}
X^{\nu}(\sigma) 
\right)_{\sigma = 0, \pi} = 0.
\label{16a} 
\end{eqnarray}
which is easily identifiable with Polyakov form of BC \cite{rb}
following from the action (\ref{Polyaction}).

\noindent 
Using phase space variables $X^{\mu}$ and
$\Pi_{\mu}$, (\ref{16}) can be rewritten as 
\begin{eqnarray}
K^{\mu} = \left[\left(\rho \Pi^{\mu} + \lambda X^{\prime \mu}\right)
+ e B^{\mu}_{\ \nu}\left(\Pi^{\nu} - eB^{\nu}_{\ \rho}
X^{\prime \rho}\right)\right]_{\sigma = 0, \pi} = 0.
\label{18} 
\end{eqnarray}
 Hence it is possible to interpret either of
(\ref{16}) or (\ref{18}) as an interpolating BC.

\noindent
Now we come to the discussion of the constraint structure
of the interpolating interacting string.
Note that the independent fields in 
(\ref{interpolatingL}) are $X^{\mu}$, $\rho$ and $\lambda$. 
The corresponding momenta denoted by $\Pi_{\mu}$, 
$\pi_{\rho}$ and $\pi_{\lambda}$, are given as:
\begin{eqnarray}
\Pi_{\mu} &=&  -\frac{1}{\lambda}\left(\dot{X}_{\mu} +
\rho X^{\prime}_{\mu}\right) + eB_{\mu\nu}X^{\prime \nu} \nonumber\\
\pi{\rho}& = & 0 \nonumber\\
\pi_{\lambda}& = & 0\,.
\label{211}
\end{eqnarray}
In addition to the PB(s) similar to (\ref{NGPB}), we now have:
\begin{eqnarray}
 \{\rho\left(\tau,\sigma\right),
 \pi_{\rho}\left(\tau,\sigma^{\prime}\right)\} =
  \delta\left(\sigma - \sigma^{\prime}\right) \nonumber\\
 \{\lambda\left(\tau,\sigma\right),
 \pi_{\lambda}\left(\tau,\sigma^{\prime}\right)\} =
  \delta\left(\sigma - \sigma^{\prime}\right).
\label{interpolatingPB}
\end{eqnarray}
The canonical Hamiltonian following from (\ref{interpolatingL}) reads:
\begin{equation}
{\cal{H}}_c = -\rho \Pi_{\mu}X^{\prime\mu} 
- \frac{\lambda}{2}\left\{\left(\Pi_{\mu} - e B_{\mu \nu}
X^{\prime \nu}\right)^2 + X^{\prime 2}\right\} 
\label{canonicalH2}
\end{equation}
which reproduces the total Hamiltonian (\ref{totalH}) of the NG action. 
From the definition of the canonical momenta we can 
easily identify the primary constraints:
\begin{eqnarray}
\Omega_{3} = \pi{\rho}& \approx & 0 \nonumber\\
\Omega_{4} = \pi_{\lambda}& \approx & 0\,.
\label{int_primary}
\end{eqnarray}
The conservation of the above primary constraints leads to the
secondary constraints $\Omega_1$ and $\Omega_2$ of (\ref{NGconstraints}).
The primary constraints of the NG action
appear as secondary constraints in this formalism\footnote{No more secondary
constraints are obtained.}. The system of constraints for the 
Interpolating Lagrangian thus comprises of the set 
(\ref{int_primary}) and (\ref{NGconstraints}). 
The PB(s) of the constraints of (\ref{int_primary}) 
vanish within themselves. Also
the PB of these with (\ref{NGconstraints}) vanish. 

\section{Modified brackets for Interpolating String}
\subsection{Free Interpolating String:}
Let us consider boundary condition for free interpolating string
which can be obtained by setting $B_{\mu \nu} = 0$ in (\ref{18}):
\begin{eqnarray}
K^{\mu} = \left[\left(\rho \Pi^{\mu} + \lambda X^{\prime \mu}\right)
 \right]_{\sigma = 0, \pi} = 0.
\label{20} 
\end{eqnarray}
It is now easy to note that the above BC is not compatible with
the basic PB (\ref{NGPB}). To incorporate this, an appropriate 
modification in the PB is in order.
In \cite{hrt, rb, rbbckk, bcsgaghs}, the equal time brackets
were given in terms
of certain combinations ($\Delta_{+}(\sigma , \sigma^{\prime})$)
of periodic delta function\footnote{The form of the
periodic delta function is given by $\delta_{P}(x-y) = \delta_{P}(x-y+ 2\pi)
=\frac{1}{2\pi} \sum_{n\in Z}e^{in(x-y)}$ and is related to the usual
Dirac $\delta$-function as $\delta_P(x-y) = 
\sum_{n\in Z}\delta(x - y + 2\pi n)$.}
\begin{eqnarray}
\{X^{\mu}(\tau , \sigma) ,  \Pi_{\nu}(\tau , \sigma^{\prime})\}
= \delta^{\mu}_{\nu} \Delta_{+}(\sigma, \sigma^{\prime})
\label{23}
\end{eqnarray}
where,
\begin{eqnarray}
\Delta_{+}\left(\sigma , \sigma^{\prime}\right)
&=& \delta_{P}(\sigma - \sigma^{\prime})
 + \delta_{P}(\sigma + \sigma^{\prime})  
= \frac{1}{\pi} + \frac{1}{\pi}\sum_{n\neq 0}
\mathrm{cos}(n\sigma^{\prime})\mathrm{cos}(n\sigma)
\nonumber \\
\Delta_{-}\left(\sigma , \sigma^{\prime}\right)
&=& \delta_{P}(\sigma - \sigma^{\prime})
 - \delta_{P}(\sigma + \sigma^{\prime}) 
=  \frac{1}{\pi}\sum_{n\neq 0}
\mathrm{sin}(n\sigma^{\prime})\mathrm{sin}(n\sigma)
\label{24}
\end{eqnarray}
 rather than an ordinary delta function to ensure 
compatibility with Neumann BC 
\begin{eqnarray}
\partial_{\sigma}X^{\mu}(\sigma)\vert_{\sigma = 0, \pi} = 0\,, 
\label{16b}
\end{eqnarray}
in the bosonic sector. Observe that the other brackets 
\begin{eqnarray}
\{X^{\mu}\left(\sigma\right), X^{\nu}\left(\sigma^{\prime}\right)\}
 & = & 0 
\label{24q}\\
\{\Pi^{\mu}\left(\sigma\right), \Pi^{\nu}
\left(\sigma^{\prime}\right)\} & = & 0
\label{24p}
\end{eqnarray}
are consistent with the Neumann boundary condition (\ref{16b}).

\noindent 
Now a simple inspection shows that the BC (\ref{20}) is also 
compatible with (\ref{23})\footnote{Note that there is no 
inconsistancy in (\ref{16b}) as
$\partial_{\sigma}\Delta_{+}\left(\sigma , \sigma^{\prime}\right)
\vert_{\sigma = 0, \pi}= 0$.} and (\ref{24p}), but not with
(\ref{interpolatingPB}) and (\ref{24q}). Hence the brackets
(\ref{interpolatingPB}) and (\ref{24q})
should be altered suitably. 

\noindent Now, since $\rho$ and $\lambda$ are odd and even
functions of $\sigma$ 
respectively, we propose:
\begin{eqnarray}
\{\rho(\tau,\sigma ) , \pi_{\rho}(\tau, \sigma^{\prime} )\} &=& 
\Delta_{-}(\sigma , \sigma^{\prime}) \nonumber \\
\{\lambda(\tau,\sigma ) , \pi_{\lambda}(\tau, \sigma^{\prime} )\} &=& 
\Delta_{+}(\sigma , \sigma^{\prime}).
\label{32j}
\end{eqnarray}
and also make the following ansatz for the bracket among 
the coordinates (\ref{24q}):
\begin{eqnarray}
\{X^{\mu}(\tau , \sigma) ,  X^{\nu}(\tau , \sigma^{\prime})\}
= C^{\mu \nu}(\sigma, \sigma^{\prime})\ \ ;\ \ 
\mathrm{where} \ \ \ C^{\mu \nu}(\sigma, \sigma^{\prime}) = 
-\  C^{\nu \mu}(\sigma^{\prime}, \sigma)\,.
\label{25}
\end{eqnarray}
One can easily check that the brackets (\ref{32j}) are
indeed compatible with the BC (\ref{20}).
Now imposing the BC (\ref{20}) on the above equation (\ref{25}), we obtain
the following condition:
\begin{eqnarray}
\partial_{\sigma}C^{\mu \nu}\left(\sigma, 
\sigma^{\prime}\right)|_{\sigma = 0, \pi} 
= \frac{\rho}{\lambda}\eta^{\mu \nu} \Delta_{+}\left(\sigma, 
\sigma^{\prime}\right)|_{\sigma = 0, \pi}\,.
\label{26}
\end{eqnarray}
Now to find a solution for 
$C^{\mu\nu}(\sigma ,\sigma^{\prime})$, 
we choose{\footnote {The condition (\ref{27}) reduces to a restricted 
class of metric for Polyakov 
formalism that satisfy $\partial_{\sigma}g_{01}=0$.
Such conditions also follow from a standard 
treatment of the light-cone gauge \cite{pol}.}}:
\begin{eqnarray}
\partial_{\sigma}\left(\frac{\rho}{\lambda}\right) = 0
\label{27}
\end{eqnarray}
which gives a solution of 
$C^{\mu\nu}(\sigma ,\sigma^{\prime})$ as:
\begin{eqnarray}
C^{\mu\nu}(\sigma ,\sigma^{\prime}) = \eta^{\mu \nu}\left[
\kappa(\sigma)\Theta(\sigma ,\sigma^{\prime}) - 
\kappa(\sigma^{\prime})\Theta(\sigma^{\prime} ,\sigma )\right] 
\label{28}
\end{eqnarray}
where the generalised step function $\Theta 
(\sigma ,\sigma^{\prime})$ satisfies,
\begin{eqnarray}
\partial_{\sigma }\Theta (\sigma ,\sigma^{\prime}) = 
\Delta_{+}(\sigma ,\sigma^{\prime})\,. 
\label{29}
\end{eqnarray}
Here, $\kappa(\sigma) = \frac{\rho}{\lambda}(\sigma)$ is a pseudo-scalar.
The $\sigma$ in the parenthesis has been included deliberately
to remind the reader that it transforms as a pseudo-scalar under 
$\sigma \to -\sigma$ and should not be read as a functional dependence.
The pseudo-scalar property of $\kappa(\sigma)$ is necessary for 
$C^{\mu\nu}(\sigma ,\sigma^{\prime})$ to be an even function of
$\sigma$ as $X(\sigma)$ is also an even function of $\sigma$
in the extended interval $[-\pi, \pi]$ of the string (\ref{37}).
\noindent
An explicit form of $\Theta(\sigma ,\sigma^{\prime})$ is given by \cite{hrt}:
\begin{eqnarray}
\Theta (\sigma ,\sigma^{\prime})={\sigma \over \pi} + {1 \over \pi }
\sum_{n\neq 0}{1\over n}\mathrm{sin}(n\sigma)\mathrm{cos}(n\sigma^{\prime})
\label{30}
\end{eqnarray}
having the properties,
\begin{eqnarray}
\Theta (\sigma ,\sigma^{\prime}) &=& 1~~~ \mathrm{for} 
~~\sigma >\sigma^{\prime}
\nonumber \\
\mathrm{and} \quad
\Theta (\sigma ,\sigma^{\prime}) &=& 0 ~~~\mathrm{for}
~~ \sigma <\sigma^{\prime}.
\label{31jj}
\end{eqnarray}
Using the above relations, the simplified structure of 
(\ref{28}) reads,
\begin{eqnarray}
\{X^\mu (\tau,\sigma ),X^{\nu}(\tau, \sigma^{\prime} )\} &=& 0
~~~\mathrm{for} ~~\sigma = \sigma^{\prime} \nonumber \\
\{X^\mu (\tau,\sigma ),X^{\nu}(\tau, \sigma^{\prime} )\} &=& 
\kappa(\sigma)\,\eta^{\mu \nu}~~~\mathrm{for} ~~\sigma >\sigma^{\prime}
\nonumber \\
&=& -\kappa(\sigma^\prime)\,\eta^{\mu \nu}~~~
\mathrm{for} ~~\sigma <\sigma^{\prime}.
\label{32}
\end{eqnarray}

\noindent
We therefore propose the brackets (\ref{23}) and (\ref{32})
as the basic PB(s) of the theory and using these
one can easily obtain the following involutive
algebra between the constraints:
\begin{eqnarray}
\{\Omega_1(\sigma) , \Omega_1(\sigma^{\prime})\} &=&  
\Omega_1(\sigma^{\prime})\partial_{\sigma}\Delta_{+}
\left(\sigma , \sigma^{\prime}\right) + \Omega_1(\sigma)
\partial_{\sigma}\Delta_{-}\left(\sigma , \sigma^{\prime}\right) \,
\nonumber \\ 
\{\Omega_1(\sigma) , \Omega_2(\sigma^{\prime})\} &=&  \left(
\Omega_2(\sigma) + \Omega_2(\sigma^{\prime})\right)
\partial_{\sigma}\Delta_{+}\left(\sigma , \sigma^{\prime}\right)\,
\nonumber \\ 
\{\Omega_2(\sigma) , \Omega_2(\sigma^{\prime})\} &=& 4 \left(
\Omega_1(\sigma)\partial_{\sigma}\Delta_{+}
\left(\sigma , \sigma^{\prime}\right) + \Omega_1(\sigma^{\prime})
\partial_{\sigma}\Delta_{-}\left(\sigma , \sigma^{\prime}\right)\right).
\label{33}
\end{eqnarray}
Note that a crucial intermediate step in the above derivation is to use the relation 
\begin{eqnarray}
\{X^{\prime\mu}(\sigma), X^{\prime\nu}(\sigma^{\prime})\} = 0 
\label{importantstep}
\end{eqnarray}
which follows from the basic bracket (\ref{32}) \cite{rb}\footnote{Note that
there were some errors in \cite{rb} and the correct constraint
algebra was given in \cite{bcsgaghs}.}. 

We now compute the algebra between the Virasoro functionals
using the modified constraint algebra (\ref{33}),
\begin{eqnarray}
\{L[f(\sigma)] , L[g(\sigma)]\} &=& L[f(\sigma)g^{\prime}(\sigma)
- f^{\prime}(\sigma)g(\sigma)].
\label{0d}
\end{eqnarray}
Interestingly, the Virasoro algebra has the same form
as that of (\ref{34a}) at the classical level.
Consequently, the alternative forms of Virasoro algebra
(\ref{34b}) is also reproduced here.

It is now interesting to observe  
that the condition (\ref{27}) (which is necessary for giving an exact NC solution (\ref{28})) reduces the gauge redundancy of the 
interpolating formalism as $\rho$ and $\lambda$ are no more independent.
Consequently, one should look for only a particular combination
of the constraints (\ref{NGconstraints}) which gives a involutive algebra.

\noindent To this end we go back to (\ref{interpolatingL})
and study the effect of 
(\ref{27}) on the free interpolating Lagrangian.
Earlier it 
contained two additional fields 
$\rho$ and $\lambda$. However the interpolating Lagrangian
depends only on one of these fields $\lambda$ (say) once
the condition (\ref{27}) is imposed and one gets the following
reduced form of the Lagrangian:
\begin{eqnarray}
\mathcal{L}_{\rm{red}} = -\frac{1}{2 \lambda}\dot{X}^{2} - 
\kappa(\sigma) \dot{X}\cdot X^{\prime}
\label{mod}
\end{eqnarray}
Owing to the condition (\ref{27}), the free canonical Hamiltonian reduces to:
\begin{equation}
{\cal{H}}_c = -\kappa\left(\sigma\right) \lambda \Pi \cdot X^{\prime} 
- \frac{\lambda}{2}\left\{\Pi^{2} + X^{\prime 2}\right\} 
\label{n1}
\end{equation}
having only one primary constraint,
\begin{equation}
\pi_{\lambda} \approx 0.
\label{n2}
\end{equation}
Conserving (\ref{n2}) with the canonical Hamiltonian (\ref{n1})
we get the secondary constraint
\begin{equation}
\Omega(\sigma) = \frac{1}{2}\left[\Pi^{2} + X^{\prime 2} + 
2\kappa\left(\sigma\right)
 \Pi\cdot X^{\prime}\right] \approx 0
\label{n3}
\end{equation}
which generates the first class algebra (in NC framework):
\begin{equation}
\left\{\Omega(\sigma),\Omega(\sigma^{\prime})\right\}  = 
2\left[\kappa\left(\sigma\right)
\Omega(\sigma) 
\partial_{\sigma}\Delta_{+}(\sigma,\sigma^{\prime}) - 
\kappa\left(\sigma^{\prime}\right) 
\Omega(\sigma^{\prime}) \partial_{\sigma^{\prime}}
\Delta_{+}(\sigma,\sigma^{\prime})\right] 
\label{n4a}
\end{equation}
We shall study the consequences of the above algebra (\ref{33}) and (\ref{n4a}) in Section 5 where we make an exhaustive analysis of gauge symmetry.

\subsection{Interacting Interpolating String:}
The Interpolating action for a bosonic string moving in
the presence of a constant background Neveu-Schwarz two-form
field $B_{\mu \nu}$ is given by,
\begin{eqnarray}
S_I= \int d\tau d\sigma \left\{ -\frac{1}{2\lambda}\left[\dot{X}^{2}+
2\rho(\dot{X}.X^{\prime})+(\rho^2-\lambda^2)X^{\prime 2}
- \lambda e\epsilon^{ab}B_{\mu\nu}
\partial_{a}X^{\mu}\partial_{b}X^{\nu}\right]\right\}
\label{40}
\end{eqnarray}
where $\epsilon^{01}=-\epsilon^{10}=+1$.
The constraint structure has already been discussed in the section 3.\\
\noindent
The boundary condition (BC) (\ref{18}) can be written in a
completely covariant form as:
\begin{eqnarray}
\left[M^{\mu}_{\ \nu}\left(\partial_{\sigma}X^{\nu}\right)
+ N^{\mu \nu}\Pi_{\nu}\right]\vert_{\sigma = 0, \pi} = 0
\label{41}
\end{eqnarray}
where,
\begin{eqnarray}
M^{\mu}_{\ \nu} &=& \left(\lambda\, \delta^{\mu}_{\nu}
- e^2  B^{\mu \rho}B_{\rho \nu}\right) \nonumber \\
N^{\mu \nu} &=& \left(\rho\, \eta^{\mu \nu} + eB^{\mu \nu}\right).
\label{42}
\end{eqnarray}
This nontrivial BC leads to a modification in the original 
(naive) PBs (\ref{NGPB}).\\
\noindent
The BC (\ref{41}) can be recast as:
\begin{eqnarray}
\left(\partial_{\sigma}X^{\mu} + \Pi_{\rho}
\left(N M^{-1}\right)^{\rho \mu}\right)\vert_{\sigma = 0, \pi} = 0.
\label{43}
\end{eqnarray}
The $\{X^{\mu}(\sigma) , \Pi^{\nu}(\sigma^{\prime})\}_{
\mathrm{PB}}$ is the same as that of the free string (\ref{23}).
We therefore make similar ansatz like (\ref{25}) and
using the BC (\ref{43}), we get:
\begin{eqnarray}
\partial_\sigma C_{\mu\nu}(\sigma ,\sigma^{\prime})
\mid_{\sigma =0,\pi} = (NM^{-1})_{\nu \mu}\Delta_+ (\sigma, \sigma^{\prime})
\mid_{\sigma =0,\pi}.
\label{44}
\end{eqnarray}
As in the free case, we restrict to the class defined by
$\partial_\sigma (NM^{-1})_{\nu \mu} = 0$ which reduces
to a restricted class of metric for Polyakov formalism.
This reproduces the correponding equation in interacting Polyakov 
string theory (see \cite{rb}, in particular Eq 52).
We therefore, obtain the following solution:
\begin{eqnarray}
C_{\mu\nu}(\sigma ,\sigma^{\prime}) &=&
{1\over 2}(NM^{-1})_{(\nu \mu )}(\sigma)\Theta (\sigma, \sigma^{\prime})
- {1\over 2}(NM^{-1})_{(\nu \mu )}(\sigma^{\prime})
\Theta (\sigma^{\prime}, \sigma )\nonumber\\
&& + {1\over 2}(NM^{-1})_{[\nu \mu]}(\sigma)[\Theta (\sigma, \sigma^{\prime}) - 1]
+ {1\over 2}(NM^{-1})_{[\nu \mu]}(\sigma^{\prime})\Theta (\sigma^{\prime}, \sigma) 
\label{45}
\end{eqnarray}
with $(NM^{-1})_{(\nu \mu )}$ the symmetric and $(NM^{-1})_{[\nu \mu]}$ 
the antisymmetric part of $(NM^{-1})_{\nu \mu}$. 

\section{Gauge symmetry}
 In this section we will discuss the gauge symmetries of the 
different actions and investigate their correspondence with the 
reparametrisation invariances. This has been done earlier for the 
free string case \cite{rbpmas1}, however the canonical symplectic 
structure for the open string were not compatible with 
the general BC(s) of the theory. 
Now we shall investigate the 
gauge symmetry with the new modified PB structures (discussed in the earlier
sections) which correctly
takes into account the BC(s) of the theory.
Importantly, the modified PB structure reveals a NC behavior 
among the string coordinates (\ref{25}, \ref{28}). As we have seen
in the previous section, an explicit
account of noncommutativity requires a gauge fixing (\ref{27}), 
thereby reducing the gauge redundancy of the interpolating 
picture. Note that the new generator of gauge transformation 
in the reduced phase space is (\ref{n3}).
For simplicity the following analysis of the gauge 
symmetry is done for the case of the free strings. 

Our discussion will be centered on the reduced interpolating 
Lagrangian (\ref{mod}) as it provides an easy access
to the analysis of gauge symmetry. The constraint structure 
of the reduced interpolating Lagrangian has already been discussed 
in the section 4.1. All the constraints are first class
and therefore generate gauge transformations on 
${\cal{L}}_{\rm{red}}$ but the number
of independent gauge parameters is equal to the number 
of independent primary first class constraints, i.e. 
one.  In the following analysis we will apply
a systematic procedure of abstracting
the most general local symmetry transformations of the Lagrangian. 
A brief review of the procedure of \cite{brr} will thus be appropriate.

    Consider a theory with first class constraints only. The set of
constraints $\Omega_{a}$ is assumed to be classified as
\begin{equation}
\left[\Omega_{a}\right] = \left[\Omega_{a_1}
                ;\Omega_{a_2}\right]
\label{215}
\end{equation}
where $a_1$ belong to the set of primary and $a_2$ to the set of
secondary constraints. The total Hamiltonian is
\begin{equation}
H_{T} = H_{c} + \Sigma\lambda^{a_1}\Omega_{a_1}
\label{216}
\end{equation}
where $H_c$ is the canonical Hamiltonian and $\lambda^{a_1}$ are Lagrange multipliers enforcing the primary constraints. The most general expression for the generator of gauge transformations is obtained according to the Dirac conjecture as
\begin{equation}
G = \Sigma \epsilon^{a}\Omega_{a}
\label{217}
\end{equation}
where $\epsilon^{a}$ are the gauge parameters, only 
$a_1$ of which are independent. By demanding the 
commutation of an arbitrary gauge variation with the 
total time derivative,(i.e. $\frac{d}{dt}
\left(\delta q \right) = \delta \left(\frac{d}{dt} q \right) $)
 we arrive at the following equations \cite{brr, htz}
\begin{equation}
\delta\lambda^{a_1} = \frac{d\epsilon^{a_1}}{dt}
                 -\epsilon^{a}\left(V_{a}^{a_1}
                 +\lambda^{b_1}C_{b_1a}^{a_1}\right)
                              \label{218}
\end{equation}
\begin{equation}
  0 = \frac{d\epsilon^{a_2}}{dt}
 -\epsilon^{a}\left(V_{a}^{a_2}
+\lambda^{b_1}C_{b_1a}^{a_2}\right)
\label{219}
\end{equation}
Here the coefficients $V_{a}^{a_{1}}$ and $C_{b_1a}^{a_1}$ are the structure
functions of the involutive algebra, defined as
\begin{eqnarray}
\{H_c,\Omega_{a}\} = V_{a}^b\Omega_{b}\nonumber\\
\{\Omega_{a},\Omega_{b}\} = C_{ab}^{c}\Omega_{c}.
\label{2110}
\end{eqnarray}
Solving (\ref{219}) it is possible to choose $a_1$ independent
gauge parameters from the set $\epsilon^{a}$ and express $G$ of
(\ref{217}) entirely in terms of them. The other set (\ref{218})
gives the gauge variations of the Lagrange multipliers.{\footnote{
 It can be shown that
these equations are not independent conditions but appear as internal
consistency conditions. In fact the conditions (\ref{218}) follow from
(\ref{219}) \cite{brr}.}}

We begin the analysis with the interpolating Lagrangian (\ref{interpolatingL}). It contains additional fields $\rho$ and $\lambda$. We shall calculate the gauge variation of these extra fields and explicitly show that they are connected to the reparametrization by a mapping between the gauge parameters and the diffeomorphism parameters. These maps will be obtained later in this section by demanding the consistency of the variations $\delta X^{\mu}$ due to gauge transformation 
 and reparametrization 
 . 

The full constraint structure of the theory comprises of the constraints (\ref{int_primary}) along with (\ref{NGconstraints}). We could proceed from these and construct the generator of gauge transformations. 
The generator of the gauge transformations of (\ref{interpolatingL})
is obtained by including the whole set of first class constraints $\Omega_{i}$ given by (\ref{int_primary}) and (\ref{NGconstraints}) as
\begin{equation}
G =\int d\sigma \alpha_{i}\Omega_{i}
\label{3112}
\end{equation}
where only two of the $\alpha_{i}$'s are the independent gauge parameters. Using (\ref{219}) the dependent gauge parameters could be eliminated. After finding the gauge generator in terms of the independent gauge parameters, the variations of the fields $X^{\mu}$, $\rho$ and $\lambda$ can be worked out. But the number of independent gauge parameters are same in both N--G (\ref{NGaction}) and interpolating (\ref{interpolatingL}) version. So the gauge generator\footnote{Note that the gauge parameters $\alpha_{1}$ and $\alpha_{2}$ are odd and even respectively under $\sigma \to -\sigma$.}
is the same for both the cases, namely:
\begin{equation}
G =\int d\sigma\left( \alpha_{1}\Omega_{1} +  \alpha_{2}\Omega_{2}\right)
\label{3112aaa}
\end{equation}
Also, looking at the intermediate first order form (\ref{subL}) it appears that the fields $X^{\mu}$ were already there in the N--G action (\ref{NGaction}). 
The other two fields of the interpolating Lagrangian are $\rho$ and $\lambda$ which are nothing but the Lagrange multipliers enforcing the first class constraints (\ref{NGconstraints}) of the N--G theory. Hence their gauge variation can be worked out from (\ref{218}). We prefer to take this alternative route.
For convenience we relabel $\rho$ and $\lambda$ by $\lambda_{1}$ and $\lambda_{2}$
\begin{equation}
\lambda_{1} = \rho \hspace{1cm} \rm{and} \hspace{1cm}
\lambda_{2} = \frac{\lambda}{2}
\label{313}
\end{equation}
and their variations are obtained from (\ref{218})
\begin{equation}
\delta\lambda_{i}\left( \sigma\right) = -  \dot \alpha_{i}
- \int d\sigma^{\prime} d\sigma^{\prime \prime} C_{kj}{}^{i}
\left(\sigma^{\prime}, \sigma^{\prime \prime}, \sigma\right)
\lambda_{k}\left(\sigma^{\prime}\right) \alpha_{j}
\left(\sigma^{\prime \prime}\right)
\label{314}
\end{equation}
where
$ C_{kj}{}^{i}
\left(\sigma^{\prime}, \sigma^{\prime \prime}, \sigma\right)$
are given by
\begin{equation}
\left\{ \Omega_{\alpha}\left(\sigma\right),\Omega_{\beta}
\left(\sigma^{\prime}\right)\right\} =  \int d\sigma^{\prime \prime}
C_{\alpha \beta}{}^{\gamma}\left(\sigma, \sigma^{\prime}, \sigma^{\prime
\prime}\right) \Omega_{\gamma}\left(\sigma^{\prime \prime}\right)
\label{315}
\end{equation}
Observe that the structure function $ V_{a}{}^{b}$ does not 
appear in (\ref{314}) since $H_{c} = 0$ for the NG theory.
The nontrivial structure functions $C_{\alpha \beta}{}^{\gamma}
\left(\sigma, \sigma^{\prime}, \sigma^{\prime \prime}\right)$ 
are obtained from the
constraint algebra (\ref{33}) as:
\begin{eqnarray}
C_{1 1}{}^{1}\left(\sigma ,\sigma^{\prime},\sigma^{\prime \prime}\right) &=&
\left(\partial_{\sigma}\Delta_{+}\left(\sigma , \sigma^{\prime}
\right)\right)\Delta_{-}\left(\sigma^{\prime} , \sigma^{\prime \prime}\right)
+ \left(\partial_{\sigma}\Delta_{-}\left(\sigma , \sigma^{\prime}
\right)\right)\Delta_{-}\left(\sigma , \sigma^{\prime \prime}\right)
\nonumber \\
C_{2 2}{}^{1}\left(\sigma, \sigma^{\prime}, \sigma^{\prime \prime}\right) &=&
4\left(\partial_{\sigma}\Delta_{+}\left(\sigma , \sigma^{\prime}
\right)\right)\Delta_{-}\left(\sigma , \sigma^{\prime \prime}\right)
+ 4\left(\partial_{\sigma}\Delta_{-}\left(\sigma , \sigma^{\prime}
\right)\right)\Delta_{-}\left(\sigma^{\prime} , \sigma^{\prime \prime}\right)
\nonumber \\
C_{1 2}{}^{2}\left(\sigma, \sigma^{\prime}, \sigma^{\prime \prime}\right) &=&
\partial_{\sigma}\Delta_{+}\left(\sigma , \sigma^{\prime}
\right)
\left[\Delta_{+}\left(\sigma , \sigma^{\prime \prime}\right) 
+ \Delta_{+}\left(\sigma^{\prime} , \sigma^{\prime \prime}\right) 
\right]
\nonumber \\
C_{2 1}{}^{2}\left(\sigma, \sigma^{\prime}, \sigma^{\prime \prime}\right) &=&
\partial_{\sigma}\Delta_{-}\left(\sigma , \sigma^{\prime}
\right)
\left[\Delta_{+}\left(\sigma , \sigma^{\prime \prime}\right) 
+ \Delta_{+}\left(\sigma^{\prime} , \sigma^{\prime \prime}\right) 
\right]
\label{319}
\end{eqnarray}
all other $ C_{\alpha b}{}^{\gamma}$'s are zero. Note that these structure functions are potentially different from those appearing in \cite{rbpmas1, rbpmas2} in the sense that here periodic delta functions are introduced to make the basic brackets compatible with the nontrivial BC.
Using the expressions of the structure functions (\ref{319})
in equation (\ref{314}) we can easily derive:
\begin{eqnarray}
\delta \lambda_{1} &=& - \dot \alpha_{1}
+ \left(\alpha_{1}\partial_{1}\lambda_{1}
 - \lambda_{1}\partial_{1}\alpha_{1} \right) 
+ 4 \left(\alpha_{2}\partial_{1}\lambda_{2} - \lambda_{2}\partial_{1}
\alpha_{2}\right)\nonumber\\
\delta \lambda_{2} &=& -\dot \alpha_{2}
+\left(\alpha_{2}\partial_{1}\lambda_{1} - \lambda_{1} \partial_{1}\alpha_{2}
\right)
+\left(\alpha_{1}\partial_{1}\lambda_{2} - \lambda_{2} \partial_{1}\alpha_{1}
\right)
\label{GM41}
\end{eqnarray}
From the correspondence (\ref{313}), we get the variations of $\rho$
and $\lambda$ as:
\begin{eqnarray}
\delta \rho &=& - \dot \alpha_{1}
+ \left(\alpha_{1}\partial_{1}\rho
 - \rho\partial_{1}\alpha_{1} \right) + 2 \left(\alpha_{2}\partial_{1}\lambda - \lambda \partial_{1}\alpha_{2}\right)
\nonumber\\
\delta \lambda &=& -2 \dot \alpha_{2}
+2 \left(\alpha_{2}\partial_{1}\rho - \rho \partial_{1}\alpha_{2}
\right)
+\left(\alpha_{1}\partial_{1}\lambda - \lambda \partial_{1}\alpha_{1}\right)
\label{GM51}
\end{eqnarray}
 In the above we have found out the full set of symmetry transformations of the fields in the interpolating Lagrangian (\ref{interpolatingL}). These symmetry transformations (\ref{GM51}) were earlier given in \cite{kaku} for the free string case. But the results were found there by inspection
\footnote{ For easy comparison identify
$ \alpha_{1} = \eta $ and $ 2 \alpha_{0} = \epsilon $ }. 
In our approach \cite{rbpmas1, rbpmas2} the appropriate transformations are 
obtained systematically by a general method applicable
to a whole class of string actions.

Now choosing 
\begin{eqnarray}\alpha_{1}(\sigma) = 2\kappa(\sigma)\alpha_{2}(\sigma)
\label{choice}
\end{eqnarray}
one can write (\ref{3112aaa}) as:
\begin{eqnarray}
G =\int d\sigma \alpha_{1}(\sigma)\left[\Pi^{2} + X^{\prime 2} + 
2\kappa\left(\sigma\right)
 \Pi\cdot X^{\prime}\right] 
\label{gen}
\end{eqnarray}
which is nothing but the generator of gauge transformation in the reduced interpolating framework (\ref{n3}). 

The nontrivial structure functions $C\left(\sigma, \sigma^{\prime}, \sigma^{\prime \prime}\right)$ obtained from
(\ref{n4a}) and (\ref{315}) are:
\begin{eqnarray}
C\left(\sigma, \sigma^{\prime}, \sigma^{\prime \prime}\right) =
2 \left[1_{\sigma}
\partial_{\sigma}\left(
\Delta_{+}\left(\sigma , \sigma^{\prime}\right)\right)
\Delta_{+}\left(\sigma , \sigma^{\prime \prime}\right) 
- 1_{\sigma^{\prime}}\partial_{\sigma^{\prime}}
\left(\Delta_{+}\left(\sigma , \sigma^{\prime}\right)\right)
\Delta_{+}\left(\sigma^{\prime} , \sigma^{\prime \prime}\right) 
\right].
\label{rev4}
\end{eqnarray}
Substituting the structure functions 
in equation (\ref{314}) yields the variation in $\lambda$ to be:
\begin{eqnarray}
\delta \lambda = - \dot \alpha
+ 2\, 1_{\sigma}\left(\alpha\partial_{\sigma}\lambda
 - \lambda\partial_{\sigma}\alpha \right).
\label{rev5}
\end{eqnarray}
which can also be obtained by substituting (\ref{choice}) in (\ref{GM51}).
We are still to investigate to what extent the exact correspondence 
between gauge symmetry and reparametrisation holds in our 
modified NC framework. This can be done very easily if we stick to the 
method discussed in \cite{rbpmas1, rbpmas2}.

To work out the mapping between the gauge parameters and the 
diffeomorphism parameters we now take up the 
Polyakov action (\ref{Polyaction}) .
Here the only dynamic fields are $X^{\mu}$.
The transformations of $X^{\mu}$  under (\ref{3112aaa})
can be worked out resulting in the following:
\begin{equation}
\delta X_{\mu}(\sigma) = \left\{X_{\mu}(\sigma), G \right\}
 = \left( \alpha_{1} X_{\mu}^{\prime}(\sigma) +  2\alpha_{2} \Pi_{\mu}(\sigma)\right)
\label{3114}
\end{equation}
We can substitute for $\Pi_{\mu}$ to obtain: 
\begin{equation}
\delta X_{\mu}
 = \left(\alpha_1 - 2\alpha_2 \, \sqrt{- g}\, g^{01}\right)X^{\prime}_{\mu}
- 2\sqrt{- g}\, g^{00}\,\alpha_2\,\dot{X}_{\mu}
\label{31141}
\end{equation}
This is the gauge variation of $X^{\mu}$ in terms of $X^{\prime}{}_{\mu}$ 
and $\dot{X}_{\mu}$ where the cofficients appear as arbitrary functions 
of $\sigma$ and $\tau$. So we can identify them with the arbitrary 
parameters  $\Lambda_{1}$ and $\Lambda_{0}$ characterising the 
infinitesimal reparametrization \cite{holt}:
\begin{eqnarray}
\tau^{\prime} & = & \tau- \Lambda_{0}\nonumber\\
\sigma^{\prime} & = & \sigma - \Lambda_{1}\nonumber\\
\delta X^{\mu} & = & \Lambda^{a} \partial_{a} X^{\mu}
= \Lambda_{0} \dot X^{\mu} + \Lambda_{1}  X^{\prime \mu}
\label{3117}
\end{eqnarray}
and that of $g_{ab}$ as:
\begin{eqnarray}
\delta g_{ab} = D_{a}\Lambda_{b} + D_{b}\Lambda_{a}
\label{3117b}
\end{eqnarray}
where
\begin{equation}
D_{a}\Lambda_{b} = \partial_{a} \Lambda_{b} - \Gamma_{ab}{}^{c} \Lambda_{c}
\label{3123}
\end{equation}
$\Gamma_{ab}{}^{c}$ being the usual Christoffel symbols \cite{holt}. 
The infinitesimal parameters $\Lambda^{a}$ characterizes reparametrisation.

Comparing (\ref{31141}) and (\ref{3117}), we get the map 
connecting the gauge parameters with the diffeomorphism parameters:
\begin{eqnarray}
\Lambda_{0} &=&  - 2\sqrt{- g}\, g^{00}\,\alpha_2 \nonumber \\
\Lambda_{1} &=& \left(\alpha_1 - 2\alpha_2 \, \sqrt{- g}\, g^{01}\right)
\label{the_map}
\end{eqnarray}
Using the definitions (\ref{rholambdaeqn}), this map can be 
cast in a better shape:
\begin{eqnarray}
\Lambda_{1} & = &  \left(\alpha_1 - 2\frac{\alpha_{2}\rho}{\lambda}
\right)\nonumber \\
\Lambda_{0} & = & - \frac{ 2 \alpha_{2}}{\lambda}
\label{themap}
\end{eqnarray}
All that remains now is to get the variation of $\rho$ and $\lambda$
induced by the reparametrisation (\ref{3117b}).

\noindent The identification (\ref{idm}) and (\ref{3117b}) 
reproduces (\ref{GM51}) as the variations of $\rho$ and $\lambda$.
This establishes complete equivalance
of the gauge transformations with the diffeomorphisms of the string.

Once again in the reduced case the condition (\ref{choice}) 
leads to the following map:
\begin{eqnarray}
\Lambda^{0}  =   - \frac{1}{\lambda}\,\alpha \quad ; \quad \Lambda^{1} = 0
\label{tm11}
\end{eqnarray}
This along with (\ref{3117b}) reproduces (\ref{rev5}) as 
the variation of $\lambda$.
The mapping (\ref{tm11}) thus establishes complete equivalance 
in the reduced case.

\section{Discussion}
In this paper, we have developed a new action formalism
for interacting bosonic string and demonstrated that
it interpolates between the NG and Polyakov form of
interacting bosonic actions. This is similar to the
interpolating action formalism for free string proposed
in \cite{rb}.
We have also modified the basic PBs in order to establish
consistency of the BC with the basic PBs. We stress that
contrary to standard approaches, BC(s) are not treated as
primary constraints of the theory. Our approach is similar in spirit
with the previous treatment of string theory \cite{hrt, rb, 
bcsgaghs}. 
The NC structures derived in our paper go over smoothly to the 
Polyakov version once suitable identifications are made. However, to give 
explicit forms of the NC structures suitable gauge fixing needs to be done.
We then set out to study the status of gauge symmetries vis-\`{a}-vis reparametrisation in this NC set up
and establish the connection between gauge symmetry and diffeomorphism
transformations. Finally, we feel that it would be interesting to investigate
whether non-critical strings can be discussed using the interpolating action in
a path-integral framework.

\section*{Acknowledgements}
Authors would like to thank Biswajit Chakraborty, 
Rabin Banerjee, Subir Ghosh and Pradip Mukherjee for some
useful discussion. AS wants to thank the Council of Scientific
and Industrial Research (CSIR), Govt. of India, for financial support
and the Director, S.N. Bose National Centre for Basic
Sciences, for providing computer  facilities. Authors would also like to thank
the referee for some useful comments and suggestions.


\end{document}